# Effect of the pseudogap on the uniform magnetic susceptibility of $Y_{1-x}Ca_xBa_2Cu_3O_{7-\delta}$

S. H. Naqib[a,b*] and J. R. Cooper[b]

[a] *Department of Physics, Rajshahi University, Raj-6205, Bangladesh*

[b] *IRC in Superconductivity and Department of Physics, University of Cambridge, Madingley Road, Cambridge CB3 0HE, UK*

**Abstract**

The effects of planar hole content, p, on the temperature-dependent uniform (q = 0) magnetic susceptibility, $\chi(T)$, of polycrystalline $Y_{1-x}Ca_xBa_2Cu_3O_{7-\delta}$ compounds with $0.0 \leq x \leq 0.20$ were investigated over a wide range of oxygen deficiency ($\delta$). Systematic variations in $\chi(T)$ were found with changing p. The characteristic pseudogap temperature scale, $E_g(p)/k_B$, was extracted from the analysis of the $\chi(T,p)$ data. We have found clear evidence that the pseudogap in the quasi-particle spectral weight appears abruptly at a planar hole content of p ~ 0.19, and that it grows rapidly with decreasing p.  © 2001 Elsevier Science. All rights reserved

## 1. Introduction

The normal and the superconducting (SC) state properties of high-$T_c$ cuprates are extremely sensitive to the number of doped holes per $CuO_2$ plane, p, and one of the most widely studied phenomena is the so-called pseudogap (PG). The PG is observed in the T-p *phase diagram* of cuprates over a certain doping range, extending from the underdoped (UD) to slightly overdoped (OD) regions. The origin of the PG is not understood yet but many of the anomalous properties can be interpreted as due to a reduction in the quasi-particle density of states (DOS) near the chemical potential [1]. Here, we report a systematic study of the uniform magnetic susceptibility, $\chi(T)$, of polycrystalline $Y_{1-x}Ca_xBa_2Cu_3O_{7-\delta}$ (Ca-Y123) over a wide range of hole content. One advantage of Ca (x) substitution is that the strongly OD region can be studied [2]. The most important information that can be extracted from the analysis of $\chi(T,p)$ is about the T- and p- dependences of the low-energy electronic density of states (EDOS) for Ca-Y123. In this study we have found clear indications that the PG energy, $E_g$, vanishes at p ~ 0.19 for Ca-Y123.

*Corresponding author. Tel.: +88-(0)721-750288; Fax: +88-(0)721-750064; e-mail: salehnaqib@yahoo.com

## 2. Experimental samples and results

High-quality polycrystalline samples of Ca-Y123 with (x = 0.0, 0.05, 0.10, and 0.20) were synthesized by solid-state reaction method using high-purity powders. The details of sample preparation and characterization can be found in refs.[3,4]. The p-values were controlled by varying both the oxygen and the Ca contents [3,4]. We have used the room temperature thermopower, *S[290K]*, as well as the quasi-universal parabolic $T_c$ - p relation [5] to determine p for all the samples. We have obtained $T_c$ from both resistivity and low-field AC susceptibility measurements. *Quantum Design* (*MPMS2* and *MPMS XL*) SQUID magnetometers were used to measure $\chi(T)$ in a field of 5 Tesla. Representative $\chi(T)$ data in Fig. 1 shows a systematic reduction in $\chi(T)$ with decreasing p for pure Y123. The same trend is seen in the 20%Ca-Y123 but only for samples with p < 0.20. For samples with p > 0.20, further overdoping does not change $\chi(T)$ in the normal state significantly, reflecting the behavior of the underlying EDOS. There is also a large Ca-dependent Curie contribution to $\chi(T)$. This rather interesting effect will be reported in detail in another paper. Here, we have subtracted the $\chi(T)$ of p = 0.22 data from the $\chi(T)$ data of other 20%Ca-Y123 samples with p < 0.22. These $\Delta\chi(T,p)$ [$\equiv \chi(T,p) - \chi(T,p = 0.22)$] data shows clearly (see Fig. 2) the changes in the EDOS with p [4].

## 3. Analysis of $\chi(T)$ data

The spin susceptibility, $\chi_s(T,p)$ represents the quasi-particle spectral density near the Fermi level, $\varepsilon_F$, i.e.,

$$\chi_s(T) = \mu_B^2 <N(\varepsilon)>_T \quad (1)$$

where, $<N(\varepsilon)>_T = \int N(\varepsilon)(\partial f/\partial \varepsilon)d\varepsilon$, is the thermal average of the EDOS, $\mu_B$ is the Bohr magneton, and f is the Fermi-function. Therefore, $\chi_s$ at any particular temperature, T, represents the average $N(\varepsilon)$ over an energy region $\sim \varepsilon_F \pm 2k_BT$. The striking, quantitative agreement between $T\chi(T)$ and the electronic entropy, obtained from electronic specific heat, is well-documented [6] and justifies interpreting $\chi(T,p)$ data in terms of the EDOS. Here, we use a V-shaped gap, $N(\varepsilon) = N_0$ for $|\varepsilon - \varepsilon_F| > E_g$ and $N(\varepsilon) = N_0|\varepsilon - \varepsilon_F|$ for $|\varepsilon - \varepsilon_F| < E_g$, giving [7]

$$<N(\varepsilon)>_T = N_0[1 - D^{-1}\ln\{\cosh(D)\}] \quad (2)$$

where, $D = E_g/2k_BT$. Eqn. (2) captures the essential features of the PG on the entropy [6]. Fig. 1a shows fits of $\chi(T,p)$ data for pure (0%Ca) Y123 samples from 400K to $\sim T_c + 30K$ (to avoid significant SC fluctuations near $T_c$) using Eqns. (1) and (2) with a p-independent value of $N_0$. $E_g(p)$ values were extracted from the fits and are shown in Fig. 3 together with the $T_c(p)$ for pure Y123.. Fig. 2 shows the $\Delta\chi(T,p)$ data for 20%Ca-Y123. As found for OD Bi2212 [7], good fits to the V-shaped gap model now require $N_0$ to fall slightly with $\varepsilon$ and are not shown. But the raw data clearly show the abrupt onset of the PG for $p < 0.2$.

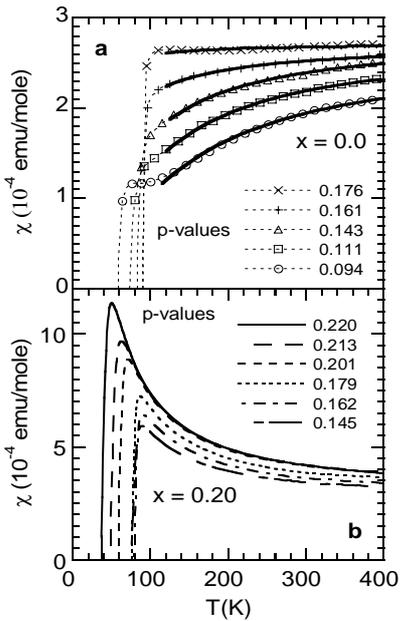

Fig. 1. $\chi(T,p)$ of (a) pure Y123 (thick full lines show the fits to Eqn. 2) and (b) 20%Ca-Y123.

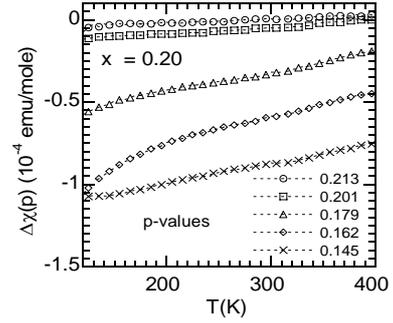

Fig. 2. $\Delta\chi(T,p)$ for 20%Ca-Y123, abrupt change occurs when p < 0.20.

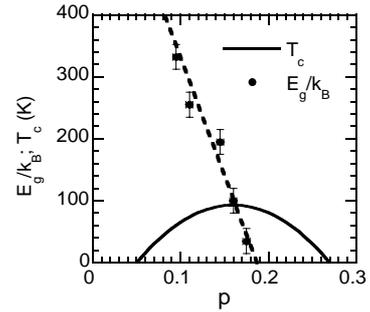

Fig. 3. $E_g(p)/k_B$ and $T_c(p)$ of pure Y123. The dashed straight line is drawn as a guide to the eyes.

## 4. Discussions and conclusions

The values of $E_g$ obtained from the analysis of the $\chi(T,p)$ data agree well with the previous reports [1,8]. For Ca doped Y123 with $p > \sim 0.19$ no sign of a PG can be detected (see Fig. 2). The results support specific heat [6] and charge transport measurements [8]. Namely, the PG coexists with superconductivity and vanishes at $p \sim 0.19$. All these point towards a non-SC origin of the PG.

### Acknowledgements

We thank Dr. J. W. Loram and Prof. J. L. Tallon for helpful comments and suggestions. SHN thanks members of the Quantum Matter group, Department of Physics, University of Cambridge, UK, for their hospitality.